\documentclass[prx,final,twocolumn]{revtex4-2}
\usepackage{bm,amsmath}
\usepackage{amssymb}
\usepackage{braket}
\usepackage[colorlinks=True,citecolor=blue,linkcolor=blue,urlcolor=blue]{hyperref}
\usepackage{graphicx}

\newcommand{\mat}[1]{\mathsf{#1}}
\DeclareMathOperator{\Tr}{Tr}

\begin{document}

\title{Fate of dissipative hierarchy of timescales in the presence of unitary dynamics}
\author{Nick D. Hartmann}
\author{Jimin L. Li}
\author{David J. Luitz}
\affiliation{Institute of Physics, University of Bonn, Nu\ss allee 12, 53115 Bonn}
\begin{abstract}
	Understanding the dynamics of open quantum systems is intrinsically important to control and utilize quantum hardware in various applications. The generic behavior of purely dissipative open quantum many-body systems with $k$-local, normal dissipation processes can be investigated using random matrix theory, revealing different decay timescales of observables determined by their complexity as shown in [Wang \emph{et al.}, \href{https://link.aps.org/doi/10.1103/PhysRevLett.124.100604}{Phys. Rev. Lett. \textbf{124}, 100604 (2020)}]. The time evolution of an open quantum system is entirely described by the eigenvalue spectrum of the Lindbladian, in which the hierarchy of decay timescales is reflected in the formation of well-separated eigenvalue clusters. Here, we extend the analysis of this spectrum to the presence of unitary dynamics using a random matrix model. In case of strong dissipation, the unitary dynamics can be treated perturbatively and the $k$-locality of the Hamiltonian determines how susceptible the spectrum is to such a perturbation. For the physically most relevant case of (dissipative) two-body interactions, the correction in the first order of the perturbation vanishes, leading to the relative robustness of the spectral features, which in turn implies the same robustness of the corresponding separation of decay timescales. We find the perturbative approach for the spectrum to yield an excellent agreement with exact diagonalization results. For weak dissipation, the spectrum flows into clusters except for outlier eigenmodes, which we identify to be the local symmetries of the Hamiltonian. We support our analytic findings by numerical simulations and show that our generic model applies to a non-random Heisenberg chain.
\end{abstract}

\maketitle

\section{Introduction}
The eigenvalues of unitary quantum many-body systems typically exhibit level repulsion. It was realized early on that while the precise location of all eigenvalues can only be analytically described in rare, integrable systems, their statistical properties in the vast majority of systems are well captured by random matrix theory \cite{wigner_random_1967,dyson_statistical_1962,dyson_statistical_1962-1,dyson_statistical_1962-2,deutsch_quantum_1991,beenakker_random-matrix_1997,wigner_random_1967,ginibre_statistical_1965,bohigas_characterization_1984}. This random matrix description provides a powerful framework, correctly predicting the distribution of gaps between neighboring eigenvalues, and further providing insight on the corresponding structure of eigenvectors and their link to thermalization \cite{srednicki_chaos_1994,rigol_thermalization_2008,dalessio_quantum_2016,deutsch_eigenstate_2018,luitz_ergodic_2017,abanin_recent_2017}.

It is exceedingly difficult to fully isolate a real quantum system from its environment, which is a source of dissipation and incoherent dynamics. Systems in contact with their environment are therefore described in the framework of open quantum systems.
In the simplest class of Markovian open quantum systems, the dynamics of the density matrix is determined by the Lindblad master equation \cite{gorini_completely_1976} and  the generator of the dynamics is a Lindbladian superoperator, rather than just the Hamiltonian. The spectrum of the Lindbladian is generally complex with nonpositive real parts, such that at long times a state from the manifold corresponding to the zero eigenvalue is reached.

Analogously to the approach for unitary quantum systems, random matrix descriptions were developed to identify universal features of Markovian open systems. 
These theories are based on an ensemble of random matrices which are at the same time valid Lindbladians and hence generate completely positive trace preserving maps. The idea is that this ensemble should exhibit \emph{generic} behavior of Lindbladians. It was found that non-local, purely dissipative Liouvillians exhibit a lemon-shaped support on the complex plane \cite{denisov_universal_2019,sa_spectral_2020,tarnowski_random_2021,can_random_2019,lange_random-matrix_2021}, which is similar to the spindle-shaped spectral support found in classical random Master equations \cite{timm_random_2009}.

With a non-vanishing Hamiltonian part, the spectral support is deformed to an elliptic shape  \cite{denisov_universal_2019}, and various spectral transitions in the weak and strong dissipation limits \cite{can_spectral_2019,haga_quasiparticles_2022} and transitions of the steady-state properties \cite{sa_spectral_2020,costa_spectral_2022} were identified. 

To bridge the gap between these purely random matrix theories and physical systems, the $k$-locality of (dissipative) interactions in a complexity theoretic sense was included. The central idea of this new ensemble of Lindbladians is to investigate the \emph{generic} properties of markovian open quantum systems with \emph{few-body} interactions. 
In particular, the most physically relevant case of two body interactions leads to a random matrix ensemble with an additional striking structure:
Instead of one lemon-shaped eigenvalue cluster, well-separated clusters of eigenvalues were identified, which correspond to observables of specific $k$-locality. Hence in the presence of dissipative few-body interactions, $k$-local observables generically relax to the stationary state at different rates which are determined only by $k$ \cite{wang_hierarchy_2020}. This phenomenological prediction was recently verified in experiments on a noisy qubit platform \cite{sommer_many-body_2021} and extended to include the case of strongly varying dissipation strengths, which introduce additional metastable structure \cite{li_random_2022}. Such setup is commonly encountered in physical problems, especially in active questions over the past years, such as transport properties of open quantum spin chains~\cite{benenti_negative_2009, benenti_charge_2009, nava_lindblad_2021}, slow quantum dynamics~\cite{macieszczak_towards_2016, rose_metastability_2016} and classical models through Lindbladian mappings~\cite{leeuw_constructing_2021, nava_traffic_2022}.

Moreover, in a recent series of works, the related problem of generic, open systems of Majorana fermions with interactions of Sachdev-Ye-Kitaev (SYK) type was considered \cite{sa_lindbladian_2022,kulkarni_lindbladian_2022,kawabata_dynamical_2022,garcia-garcia_keldysh_2023}. Interestingly, depending on the choice of dissipation processes described by either (linear combinations) of single Majorana operators or quadratic terms, different shapes of the Lindbladian spectrum was observed. For linear jump operators, the spectrum decomposes into eigenvalue clusters \cite{kulkarni_lindbladian_2022,garcia-garcia_keldysh_2023}, while for quadratic jump operators a spectrum consistent with the ``lemon'' shape of nonlocal Lindbladians is found \cite{sa_lindbladian_2022,kulkarni_lindbladian_2022}. These results are consistent with our findings because quadratic fermionic operators are non-local due to Jordan-Wigner strings connecting the sites where the fermion operators act.

In past studies, unstructured and $k$-local purely dissipative Lindbladians have been the central object. However, almost all physical processes include a component of unitary time evolution, usually represented by a Hamiltonian. Thus, it is crucial to understand the features of random $k$-local Lindbladians and the relaxation of $k$-local observables for modeling a generic open quantum system. In general, the presence of a Hamiltonian tends to merge the eigenvalue clusters observed in the purely dissipative case, as shown concretely in a dissipative Bose-Hubbard model~\cite{zhou_renyi_2021}. 
    While the separation of eigenvalues in their imaginary part is directly related to the observation of a separation of decay timescales in the system, the loss of eigenvalue cluster formation smears out this separation of timescales. In this paper, we investigate, to which extent the separation of timescales can persist in the presence of unitary dynamics by focusing on the eigenvalue spectrum of the Lindbladian.
We study the interplay of the competing components: unitary and dissipative parts of random Lindbladians with two-body (i.e. $k=2$ local) dissipative interactions.  

By treating the Hamiltonian as a perturbation, we show how the purely dissipative eigenvalue clusters deform into a single stripe, corresponding to a loss of separation of timescales, if the Hamiltonian contribution is significant. 
The predicted averaged eigenvalue shows a good agreement with the exact diagonalization results. In addition, we discover outlier eigenmodes which survive away from the strip when the dissipation is weak; This means that some dissipative timescales remain well separated from the rest. Lastly, we fix the Hamiltonian to be the non-random Heisenberg model and demonstrate that our results for the generic case are valid in this specific model.

\section{Model}
For Markovian open quantum systems, the Liouvillian superoperator $\mathcal{L}$ generates the dynamics of density matrices and the Lindblad master equation
reads
\begin{align}
\frac{d \rho}{dt} &=  \mathcal{L}(\rho) =  \alpha \mathcal{L}_{U}(\rho) + \mathcal{L}_{D}(\rho) \\
 &= -i \alpha [H,\rho] + \sum_{n,m=1}^{N_{L}}K_{nm}\left[ L_{n} \rho L_{m}^{\dagger} - \frac{1}{2}\{ L_{m}^{\dagger} L_{n} , \rho \} \right] 
,
\label{eq:Lindblad_master}
\end{align}
where $H$ is the Hamiltonian yielding unitary evolution by $\mathcal{L}_U$ and $\{L_{n}\}_{n=1,\dots, N_L}$ is the set of $N_{L}$ Lindblad (jump) operators that represent the dissipation channels coupled by the positive semidefinite Kossakowski matrix $K$ \cite{breuer_theory_2002}. By choosing $K$ as a random matrix we aim to model a generic system. $1/\alpha\ge 0$ parametrizes the relative strength of dissipation.

For a generic modeling of dissipative quantum many-body spin systems, we consider $\ell$ spins $S=1/2$ with a Hilbert space dimension $N = 2^{\ell}$ and a Liouville operator space dimension $N^{2}=4^{\ell}$. The $k$-locality of dissipative interactions is introduced by choosing the Lindblad operators to be normalized Pauli strings 
\begin{equation}
L_{\vec{\mu}} = \frac{1}{\sqrt{N}} \sigma_{\mu_1} \otimes \sigma_{\mu_2} \otimes \cdots \otimes \sigma_{\mu_\ell}, \quad \mu_i \in \{0,x,y,z\},
\label{eq:Pauli_strings}
\end{equation}
of maximal weight $k_{\vec{\mu}} = \sum_{i=1}^{\ell} (1 - \delta_{\mu_i 0 }) \leq k_\text{max}$, i.e. there are at most $k_{\text{max}}$ non-identity operators in the string. We note that Pauli strings with weight $k$ can be interpreted as $k$-body interactions and we refer to weight $k$ strings as $k$-local operators. It is important to emphasize that $k$-locality is not connected to a specific geometry of the system and more general than spatial locality.
These used jump operators are traceless $\Tr(L_{n}) = 0$ and satisfy orthonormality $\Tr( L^{\dagger}_{n}L_{m} ) = \delta_{n,m}$ \cite{kitaev_classical_2002}.

We consider one and two-body dissipation channels ($k_\text{max} = 2$), so the total number of traceless Lindblad operators is $N_L= 3 \ell + 9 \ell (\ell - 1) /2$ and as explained we choose the Kossakowski matrix $K$ to be a random matrix. 
To ensure that $K$ is positive semidefinite, we first sample a diagonal matrix $D$ with i.i.d. non-negative entries from a uniform distribution. The matrix $D$ is normalized by $\mathrm{Tr}D = N$, and then rotated into a random unitary basis by a unitary transformation $U$ sampled from the Haar measure to yield $K = U^{\dagger}DU$. 
The mean of the off-diagonal elements vanishes, $\mathrm{mean}(K_{ij})=0$ with a standard deviation $\mathrm{std}(K_{ij})=d/\sqrt{6N_L}$, where $d$ is the mean of the diagonal elements $\mathrm{mean}(K_{ii}) = N / N_L = d$. Consequently, $K$ is diagonally dominant and can therefore be separated into a large diagonal term and a small off-diagonal term, which is the basis for the perturbation theory proposed in Ref. \cite{wang_hierarchy_2020}.

The effect of $U$ is to rotate the (hermitian) Pauli string operators $L_n$ into non-hermitian Lindblad operators, since the coefficients of $U$ are generally complex. The transformed Lindblad operators $L_{\nu} = \sum_n U_{\nu n} L_n$ in the diagonal basis of the Kossakowski matrix are however normal, since they commute with their adjoint operators, $[L_{\nu}^\dagger , L_{\nu}]=0$. The ensemble hence consists of $k$-local, normal Lindblad operators.

For the unitary dynamics we focus on 2-body interactions, in the form of all possible Pauli strings of weight $k=2$ \cite{sachdev_gapless_1993}, and start from a generic Hamiltonian with all-to-all interactions
\begin{equation}
H = \sum_{s,s' = \{x,y,z\} } \sum^{\ell}_{i < j } J^{s,s'}_{i,j}  \sigma^{s}_{i} \sigma^{s'}_{j},
\label{eq:Hamiltonian_random}
\end{equation}
where the random real coupling constants $J^{s,s'}_{i,j}$ follow independent Gaussian distributions centered at 0. The notation $\sigma^s_i = I\otimes \sigma^s \otimes I$ is the weight one Pauli string acting on site $i$. We normalize the Hamiltonian by $\Tr(H^{2}) = N$ and with the following standard deviation 
\begin{equation}
    \sigma = \mathrm{std}(J^{s,s'}_{i,j}) = \sqrt{ 2/ (9 \ell(\ell -1))}.
    \label{eq:std_J}
\end{equation}

Before we provide our analytic results in full detail, it is worth outlining the strategy of our analysis.
Our starting point is the purely dissipative case $\alpha=0$, for which the spectrum of the Liouvillian $\mathcal L$ splits into well-separated eigenvalue clusters. We consider a three term decomposition of $\mathcal L$ into $\mathcal L_D^0 + \mathcal L_D^1 +\alpha \mathcal L_U$ and the key insight is that $\mathcal L_D^0$ is diagonal in the basis of Pauli strings. For small $\alpha$, we can treat the unitary part $\alpha \mathcal L_U$ as a perturbation and analyze the deformation of the spectrum. In the last step, we include the off-diagonal part of the dissipator $\mathcal L_D^1$, which leads to further corrections in the spectrum but does not change the physical structure of the eigenmodes significantly.

\section{Spectrum}
\begin{center}
\begin{figure*}[hbt!]
    \centering
    \includegraphics[width=\textwidth]{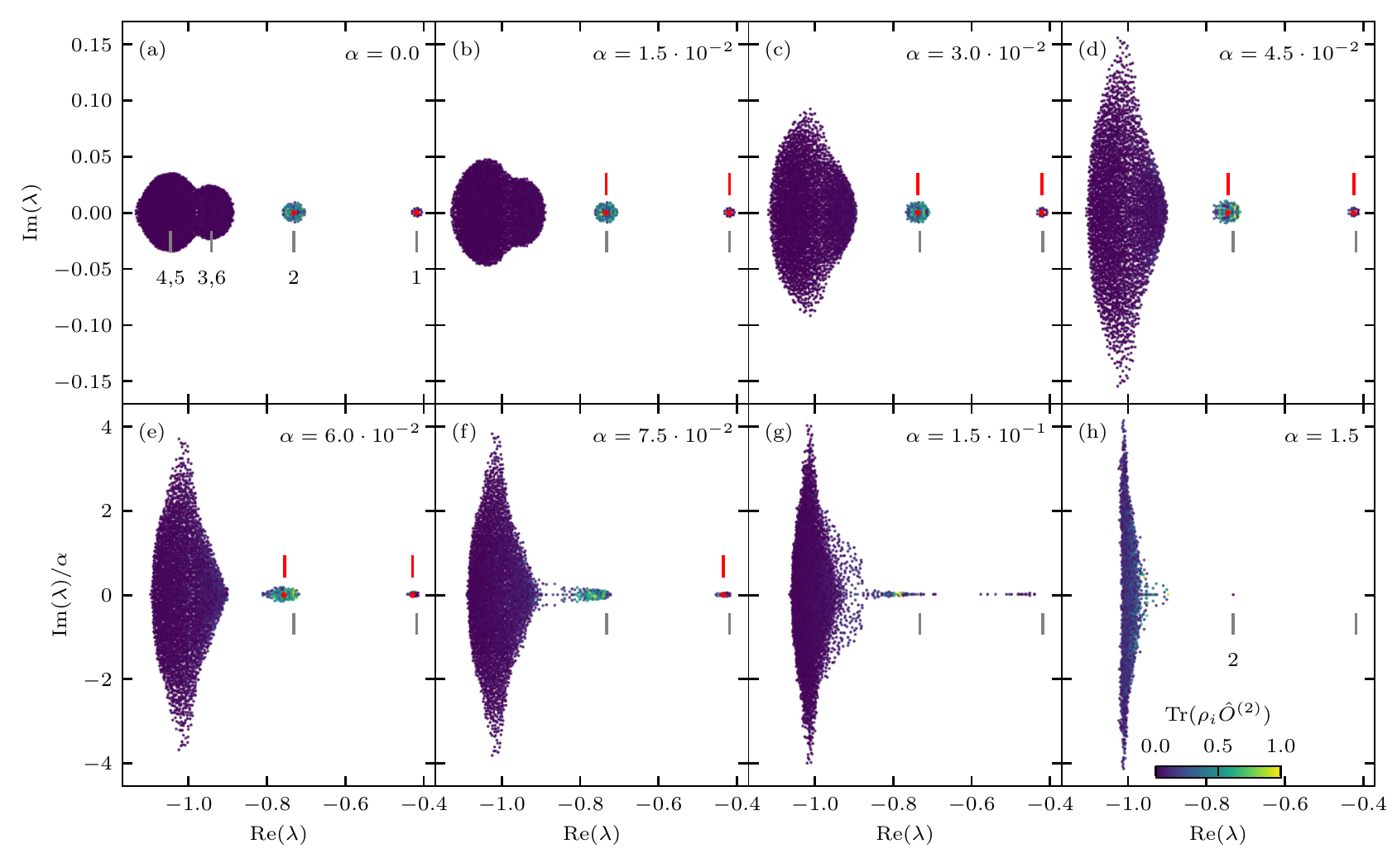}
    \caption{Eigenvalue spectrum of one realization of the random Liouvillian for a system size $\ell = 6$, with weight-2 $\mathcal{L}_{U}$ and up-to-weight-2 ($k\leq 2$) $\mathcal{L}_{D}$ at eight different relative strengths $\alpha$ of the unitary component. 
    Note that in the bottom panels we rescale the imaginary part of the eigenvalues by a factor of $1/\alpha$ to put these panels on the same scale. The color indicates the overlap $\mathrm{Tr}(\rho_i \Hat{O}^{(2)})$ of the eigenmodes $\rho_i$ with a random weight-2 operator $\Hat{O}^{(2)}$, indicating the two-body operator content of the eigenmodes. The gray lines below the spectrum give the analytic predictions for the cluster centers using $\mathcal{L}^{0}_{D}$ \cite{wang_hierarchy_2020}.
        (a) Purely dissipative case ($\alpha = 0$), the spectrum consists of separated eigenvalue clusters that are supported by weight $k$ Pauli strings.
        (b)--(g) Increasing the contribution of unitary dynamics, the clusters merge into each other and show an expansion in the imaginary direction. The red lines above the spectrum indicate the prediction of the shifted cluster centers $\lambda_0(k) + \langle \overline{\lambda_{2i}}(k) \rangle \alpha^2$ by perturbation theory, and the red dots indicate the mean of the still separated clusters.
    (h) For strong unitary dynamics, the eigenvalues are concentrated around $\mathrm{Re}(\lambda) = -1$ with a separated eigenvalue near the eigenvalue $\lambda_0(k=2)$ of $\mathcal{L}_D^0$ (gray line marked with 2).}
    \label{fig:spectrum_random}
\end{figure*}
\end{center}
For each random realization, we numerically found that the Liouvillian does not consist of any non-trivial Jordan blocks. Thus, the time evolution $\rho(t) = \mathrm{e}^{\mathcal{L}t} \rho_0$ of an initial state $\rho_0$ can be decomposed in terms of a set of bi-orthogonal right/left-eigenmodes $\rho^{R/L}_i$ (with eigenvalue $\lambda_i$) of $\mathcal{L}$, which contribute with coefficients $\mathrm{Tr}(\rho^{L}_i \rho_0) \mathrm{e}^{\lambda_i t}$. Hence the spectrum of $\mathcal{L}$ is closely related to the real-time dynamics of an arbitrary initial state; $\mathrm{Re}(\lambda_i)$ represents the relaxation timescale of the corresponding eigenmode $\rho_i$, and $\mathrm{Im}(\lambda_i)$ gives the coherent oscillation timescales.

Figure \ref{fig:spectrum_random} shows the eigenvalues of the Liouvillian $\alpha \mathcal{L}_U + \mathcal{L}_D$ from one random realization of $J$ and $K$ for different relative strengths $\alpha$ of the unitary dynamics. The spectrum is complex with a non-positive real part such that the continuous dynamical map is completely positive and trace preserving (CPTP) \cite{lindblad_generators_1976}.

For $\alpha = 0$, the dynamics is purely dissipative and eigenvalues are organized in well-separated clusters representing separated relaxation timescales, which result from the locality of dissipation \cite{wang_hierarchy_2020}. The spectral structure can be unraveled by the following arguments: the diagonally dominant nature of $K$ allows us to split the dissipator into two terms 
\begin{equation}
    \mathcal{L}_{D} = \mathcal{L}^{0}_{D} + \mathcal{L}^{1}_{D},
\label{eq:dissipator_split}
\end{equation}
where $\mathcal{L}^{0}_{D}$ contains the diagonal part with Kossakowski matrix $K_0 = d \mathbf{1}$ and and $\mathcal{L}^{1}_{D}$ the off-diagonal perturbation $K^\prime=K-d\mathbf{1}$ correspondingly. The first term is diagonal in the Pauli string basis, and the eigenvalues are determined only by the weight $k$ of the eigenvectors, which leads to highly degenerate eigenvalues $\lambda_0(k)$ of $\mathcal{L}^{0}_{D}$. Degenerate perturbation theory with $\mathcal{L}^{1}_{D}$ in each degenerate subspace lifts this degeneracy and eigenvalue clusters emerge, centered around $\lambda_0(k)$. Combinatorics of commuting operators in Eq. (\ref{eq:Lindblad_master}) give an analytic expression for the cluster centers $\lambda_0(k) = -2(6 k \ell - 4 k^2)/N_L$ \cite{wang_hierarchy_2020}.

\begin{figure}
	\centering
	\includegraphics[]{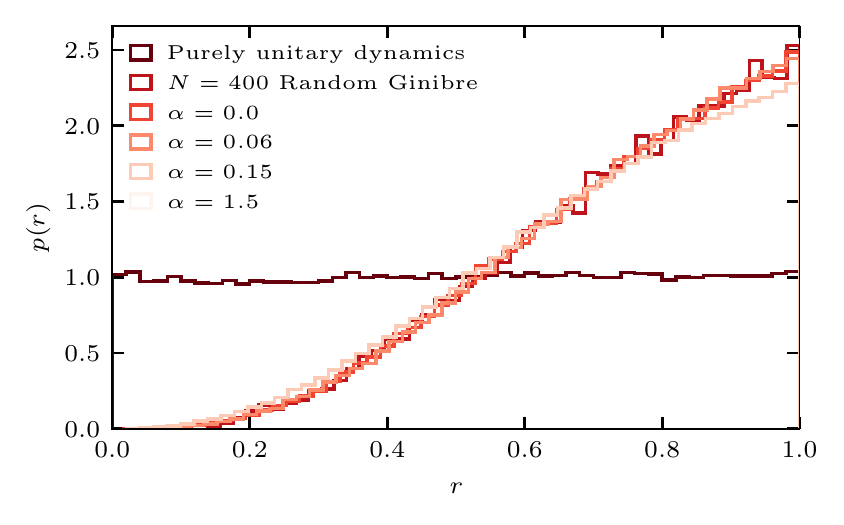}
	\caption{Probability density of the complex spacing ratio $r = |\lambda_{0} - \lambda_\text{nn}| / |\lambda_{0} - \lambda_\text{nnn}|$ for spectra as shown in Fig. \ref{fig:spectrum_random} for $\ell = 6$ from 100 realizations. Due to the symmetry with respect to the real axis, we only consider eigenvalues $\lambda$ with $\mathrm{Re}(\lambda) > 0$. The complex spacing ratio for all four unitary strengths shows clear level repulsion and agrees with the Ginibre ensemble. The purely unitary case is also considered for comparison. In contrast to the level repulsion in chaotic open systems, here $p(r) \approx 1$ matches the prediction $p(r)=1$ for unitary Poisson statistics \cite{sa_complex_2020}.}
	\label{fig:level_spacing}
\end{figure}

Once the unitary component is added ($\alpha>0$), we observe an attraction of the eigenvalue clusters along the real axes and the expansion of cluster height in the imaginary direction \cite{popkov_full_2021}. A similar declustering effect is also found in a non-random dissipative Bose-Hubbard model \cite{zhou_renyi_2021}. We observe a linear scaling of the imaginary cluster extent with $\alpha$ upon adding a local Hamiltonian, which also holds for a non-local Hamiltonian \cite{denisov_universal_2019}. Gradually increasing the unitary strength $\alpha$ leads to clusters merging when their boundaries touch. Those observations are stable against the random realization of Kossakowski matrix. Appendix \ref{appendix:statistics} shows the spectral density, also known as the density of states, for 100 realizations to demonstrate the self-averaging property. Fig. \ref{fig:level_spacing} shows the complex spacing ratio of the merged cluster. For all $\alpha$, the spectral statistics show level repulsion, which behaves like the Ginibre ensemble; even for a large unitary component $\alpha = 1.5$.

The spectral influence of the unitary term also appears in the support of the eigenmodes. For $\alpha = 0$, the support of eigenmodes is well approximated by the degenerate subspaces of $\mathcal{L}^{0}_{D}$, which are spanned by all Pauli strings with fixed weight $k$. The degeneracy of these eigenvalues is lifted by $\mathcal{L}^{1}_{D}$, leading to eigenvalue clusters with eigenmodes composed of fixed weight Pauli strings.

For $\alpha > 0$, these clusters tend to merge and the eigenmodes become smeared out over Pauli strings of different weights. We demonstrate this change of the support of eigenmodes by considering the overlap between every left eigenmode $\rho_i$ of the Liouvillian with an operator $\Hat{O}^{(2)}$ given by a random superposition of Pauli strings of fixed weight $k=2$. This is shown in the colorbar of Fig. \ref{fig:spectrum_random}. For $\alpha=0$, the eigenvalue cluster labeled $2$ has a strong weight-2 operator support. Upon increasing $\alpha$, the weight-2 nature gets diluted as Pauli strings from different degenerate subspaces start contributing.

In the large $\alpha$ limit, all clusters merge with the real part concentrating around eigenvalue $-1$ except a persistent decay mode constructed by Pauli strings of weight 2. This outlier, labeled by the gray bar in Fig. \ref{fig:spectrum_random}(h), is a unique feature of unitary locality and is absent in random ensembles \cite{sa_spectral_2020}.

\subsection{Strong dissipation: Perturbation Theory}\begin{figure*}
    \centering
    \includegraphics[width = \textwidth]{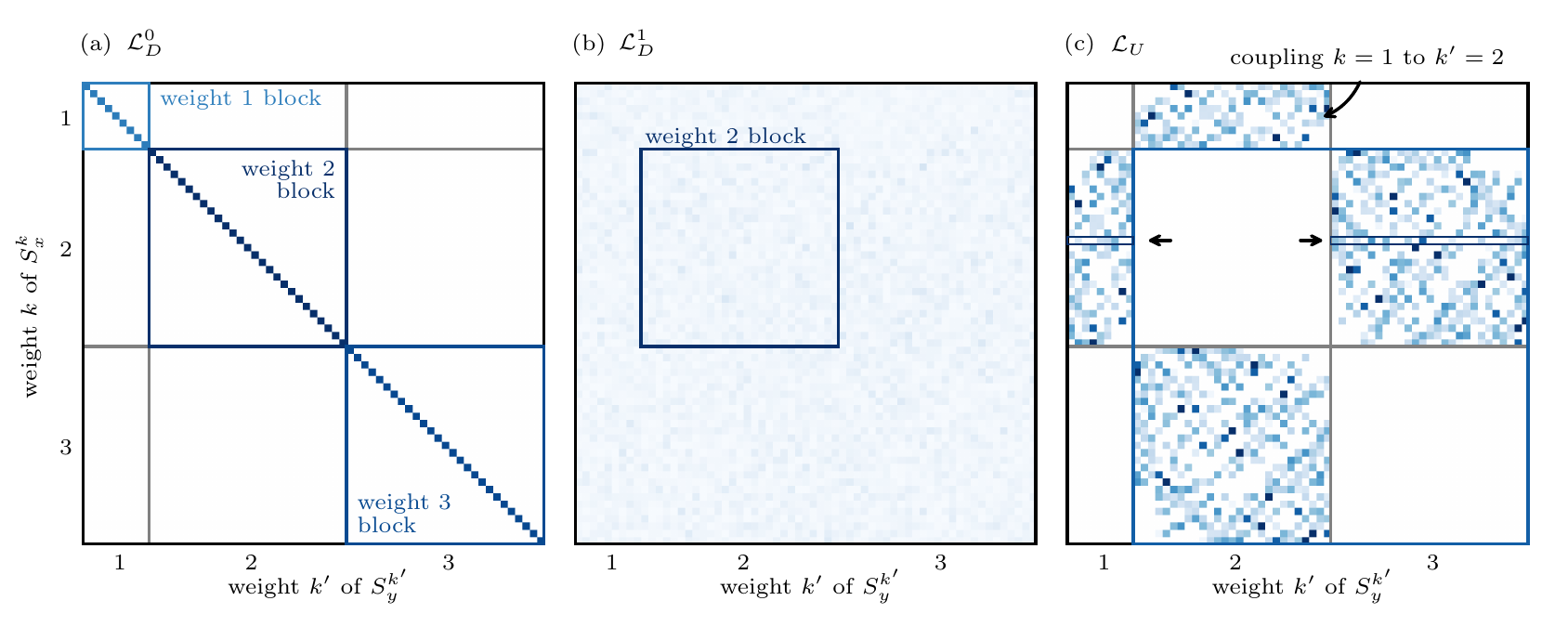}
    \caption{Single realization of the supermatrix elements of $\mathcal{L} = \mathcal{L}^{0}_{D} + \mathcal{L}^{1}_{D} + \mathcal{L}_{U}$ in the Pauli String basis, where the shade indicates the absolute value of the matrix elements. (a) and (b) show the diagonal dominant nature of $\mathcal{L}_{D}$ with eigenvalues that only depend in first order ($\mathcal{L}^{0}_{D}$) only on the Pauli weight. (c) shows a block-subdiagonal structure defined by Eq. (\ref{eq:LU_PT}). By combinatorics (cf. Eq. (\ref{eq:h_k_k_prime})) it is possible to count the number of non-vanishing elements in the rows (one row indicated by arrows). The lower right blue frame shows exemplary how the matrix $\mat{L}^{U (2)}_{k_{-},k_{+}}$ (cf. Eq. (\ref{eq:lu_adjacent})) can look like for adjacent weights $k_{-},k_{+}$ (here $2,3$).}
    \label{fig:fig2}
\end{figure*}

In this section, we provide a quantitative understanding of the observed cluster attraction by treating the unitary term $\alpha \mathcal{L}_{U}$ as a perturbation for small $\alpha$ to an otherwise purely dissipative Liouvillian $\mathcal{L}_{D}$ ($\alpha=0$). We observe a block structure of a relevant matrix, explain how this emerges from the chosen locality of $H$ by a case differentiation and use the comprehension of this structure to apply degenerate perturbation theory.

For $\alpha =0$, we start from $\mathcal{L}_{D}^{0}$ in Eq. \eqref{eq:dissipator_split}, which is diagonal in the Pauli string basis due to the CPTP condition of the Liouvillian superoperator. Furthermore, the diagonal elements of $\mathcal{L}_D^0$ in the Pauli string basis are identical for strings of the same weight, see Fig. \ref{fig:fig2}(a). In Ref. \cite{wang_hierarchy_2020}, it was worked out that the diagonal dominant structure of $\mathcal{L}_{D}$ in the Pauli string basis leads to the separation of eigenvalue clusters for the low weights Pauli strings, which is robust in the thermodynamical limit. This suggests that discarding $\mathcal{L}_D^1$ well approximates $\mathcal{L} = \mathcal{L}_D + \alpha \mathcal{L}_U$ qualitatively. In the rest of this section, we derive analytic results of $\mathcal{L}_D^0 + \alpha \mathcal{L}_U$ to understand the spectral transition. 

Then, we treat the unitary generator $\mathcal{L}_{U}$ for a 2-local Hamiltonian as a perturbation and represent it as a matrix $\mat{L}_U$ in the orthonormal Pauli string basis, which is the eigenbasis of $\mathcal{L}_D^0$. The matrix elements are given by 
\begin{equation}
    \mat{L}^{U}_{x,y} = \Tr \bigl(  S^{k}_{x} \mathcal{L}_{U}[S^{k'}_{y}] \bigr) = i\Tr\bigl( [S^{k}_{x}, S^{k'}_{y}]H\bigr).    
\label{eq:LU_PT}
\end{equation}
where $S^{k}_{x}$ (analogously $S^{k^\prime}_{y}$) represents a weight $k$ ($k^\prime$) Pauli string indexed by a subscript $x$ ($y$) which denotes the position of this Pauli string in the basis we use.
Eq. (\ref{eq:LU_PT}) encodes how the unitary term non-trivially couples different weight sectors of the Liouvillian $\mathcal{L}_D$, and hence potentially mixes previously well-separated eigenvalue clusters.

$\mat{L}_U$ shows up in a block-subdiagonal structure with respect to the $k$-local blocks (cf. Fig. \ref{fig:fig2}). In the following section we explain this structure; for this it is crucial to understand the conditions for nonzero matrix elements between sectors of different weights $k$.

First, we note that the commutator $[S^k_x, S^{k'}_y]$ is also a Pauli string and evidently must yield a term of the Hamiltonian (any weight-2 Pauli string in this work) to give a non-zero trace and hence a non-vanishing matrix element $\mat{L}^{U}_{x,y}$. We find that $\Tr\bigl( [S^{k}_{x}, S^{k'}_{y}] H\bigr) \neq 0$ is only possible if $|k - k'| = 1$, i.e., the weight of two strings differs by one. To see this, we consider two Pauli strings with all possible ranges of $|k - k'|$.

For any two Pauli strings $S^k_x$ and $S^{k'}_y$ with $|k - k'| \ge 2$, there are at least two sites of $S^{k}_{x}$ containing non-identity Pauli strings where $S^{k'}_{y}$ is identity. The commutator of the remaining sites cannot yield an identity Pauli string since this would contradict $\mathrm{Tr}([A,B]) = \mathrm{Tr}(AB)-\mathrm{Tr}(BA)=0$. Consequently, the commutator $[S^{k}_{x}, S^{k'}_{y}]$ either vanishes or has a weight higher than the 2-local Hamiltonian, hence yielding a zero element $\mat{L}^U_{x,y}$.

For $k = k'$, we partition two strings into the following three parts: 1. $n_{1}$ different non-identity Pauli matrices on the same sites, 2. a staggered arrangement of non-identity Pauli matrices and identities on $n_{2}$ sites, and 3. the same Pauli matrices on the remaining sites with $n_{3}$ non-identity Pauli matrices:
\begin{align}\label{eq:strings}
            S^{k}_{x} &= A \otimes \sigma_i \otimes \cdots \otimes \sigma_j \otimes I \otimes \cdots \otimes I \otimes \sigma_\alpha \otimes \cdots \otimes \sigma_\gamma \nonumber\\
            S^{k'}_{y} &= A \otimes \underbrace{I \otimes \cdots \otimes I  \otimes \sigma_k \otimes \cdots \otimes \sigma_l}_\text{$n_2$ sites} \otimes \underbrace{\sigma_\beta \otimes \cdots \otimes \sigma_\delta}_\text{$n_1$ sites}\nonumber\\
            A &= I \otimes \cdots \otimes I \otimes \underbrace{\sigma_n \otimes \cdots \otimes \sigma_m}_\text{$n_3$ sites}.
    \end{align}
Counting all non-identity Pauli matrices $2 n_1 + n_2 + 2 n_3 = 2k$ in both strings shows that $n_{2}$ is even. 
Evaluating the commutator of string pairs with non-zero $n_{2}$ gives either $0$ or a weight higher than two, requiring a higher locality Hamiltonian to produce non-zero matrix elements. For $n_{2}=0$, the commutator yields either $0$ or a string with odd weight, which is orthogonal to the Hamiltonian that contains only terms of weight two.

Therefore, a 2-local Hamiltonian couples a weight $k$ sector only to the $k \pm 1$ sectors of the Pauli string basis, i.e. $\mat{L}^{U}_{x,y}$ is block-subdiagonal as shown in Fig. \ref{fig:fig2}(c). The non-zero matrix elements are given by $\mat{L}^{U}_{x,y} = 2 J^{s,s'}_{i,j}$, where $(s,s',i<j)$ encode the Pauli string of weight 2 obtained by $[S^{k}_{x}, S^{k'}_{y}] = 2 \sigma^{s}_{i} \sigma^{s'}_{j}/N$. Using combinatorics, we observe that the numbers of non-zero matrix elements are equal for different rows within the same weight block; for a fixed $S^{k}_{x}$ and weight $k^\prime = k \pm 1$, it is
\begin{align} \label{eq:h_k_k_prime}
	\nonumber h(k, k^\prime=k+1)&=6k(\ell -k),\\
	h(k,k^\prime=k-1)&=2k(k-1).
\end{align} 


Focusing on $\mathcal L_D^0 + \mathcal L_U$, we obtain analytic predictions for the centers of eigenvalue clusters, and the role of $\mathcal L_U$ is to couple Pauli strings with different weights $k$ and $k\pm1$. 
In the last step, we include $\mathcal{L}_D^1$, introducing further renormalization of the resulting eigenvalues. This results in the scattering of the full spectrum around the analytic predictions for the centers of eigenvalue clusters.

To obtain a quantitative understanding, we expand the eigenvalues up to the second order in the strength $\alpha$ of the perturbation $\lambda_i = \lambda_{0i} + \lambda_{1i}  \alpha + \lambda_{2i} \alpha^2 + \mathcal{O}(\alpha^3)$. For a degenerate $\lambda_{0i}$, the first-order correction is given by diagonalizing $\mat{L}^{U}$ in the respective degenerate sub-block. Using the fact that $\mat{L}^{U \dagger} = -\mat{L}^{U}$, the first-order perturbation is purely imaginary. 

We note that while the eigenvalues of $\mathcal L_D^0$ are determined by the weight of the Pauli string eigenmodes, two subspaces of different weights can have the same eigenvalue, which results in a cluster supported by two weights.
However, for clusters supported by only a single weight, the degenerate sub-block is given by a single diagonal block in the block-structure of $\mat{L}^{U}$ and we do not expect a first-order correction, i.e. a linear scaling of the cluster height, as a consequence of vanishing diagonal blocks, as observed in Fig. \ref{fig:fig2}(c). This also holds for clusters supported by a degenerate subspace that contains weights differing by more than one. The exception are weights $k_\pm$ which are consecutive and share the same eigenvalue $\lambda_0(k_\pm)$, in our specific model $k_\pm = 3\ell /4 \pm 1/2$ for $\ell \equiv 2 \mod 4$. In this case, we expect a strong response of the spectrum to the perturbation $\mathcal L_U$.

In the following paragraph, we sketch a procedure to quantify this first-order response of the spectrum. In perturbation theory, we have to diagonalize $\mat{L}^U$ in the (degenerate) union of the $k_+$ and $k_-$-local operator subspaces, 
\begin{align} \label{eq:lu_adjacent}
    \mat{L}^{U (2)}_{k_{-},k_{+}}=\begin{pmatrix}
        0 & \mat{V} \\
        -\mat{V}^T & 0
    \end{pmatrix}
\end{align}
where $\mat{V}$ is a real matrix of matrix dimension $n_{k_-} \times n_{k_+}$, and $n_k := \binom{\ell}{k} 3^k$ is the number of Pauli strings with weight $k$. The first-order perturbation is then given by the imaginary eigenvalues $\lambda$ of $\mat{L}^{U (2)}_{k_{-},k_{+}}$.
We average over the cluster and calculate the mean of the squared modulus of the eigenvalues $\overline{|\lambda|^2}$. For a square matrix $\mat{M}_{N\times N}$, this is related to the matrix elements via $\overline{|\lambda|^2} = \frac{1}{N} \mathrm{Tr}(\mat{M}^\dagger \mat{M}) = \frac{1}{N} \sum_{i,j} |\mat{M}_{ij}|^2$. One obtains an analytic prediction by further Gaussian averaging this formula and using the knowledge of the structure of $\mat{M} = \mat{L}^{U (2)}_{k_{-},k_{+}}$ (cf. Eq. (\ref{eq:h_k_k_prime})). However, with our model it is not possible to observe the weight-$k_\pm$ cluster independently, since its separation to the other clusters is already broken by $\mathcal L_D^1$.

Still, as shown in Fig. \ref{fig:spectrum_random}, only the eigenvalue cluster with the largest negative real part ($k=3 \text{ to } 6$) shows a clear linear perturbation along the imaginary axis, since the weight-4 and 5 Pauli strings belong to the same degenerate subspace. So far, we have shown how the imaginary parts of the purely dissipative spectrum respond to the unitary term in first-order perturbation. The 2-local structure of the Hamiltonian only couples clusters corresponding to adjacent weights, which is indeed observed in Fig. \ref{fig:spectrum_random}.

\begin{figure}[]
	\centering
	\includegraphics[]{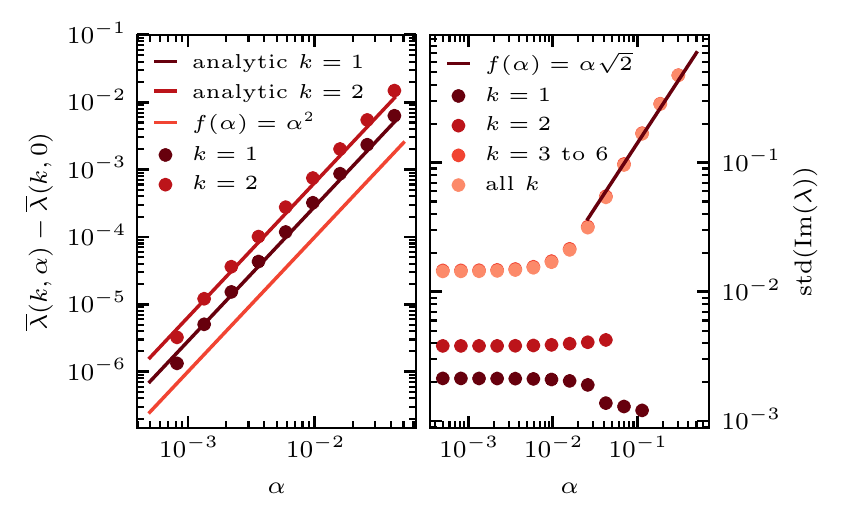}
	\caption{Numerical calculation of the influence of unitary dynamics for different relative dissipation strength $1/ \alpha$ for one realization of $\mathcal{L}_U$ and $\mathcal{L}_D$. (a) dots: numerical calculation of the shift of the cluster centers $\overline{\lambda}(k, \alpha) - \overline{\lambda}(k, 0)$, lines: analytic predictions given in Eq. (\ref{eq:second_order_PT_averaged}). The deviation of the numerical results from the analytic results is expected due to $\mathcal{L}_D^1$. The shift of the cluster centers is clearly second order in $\alpha$. (b) shows the standard deviation $\mathrm{std}(\mathrm{Im}(\lambda))$ of the eigenvalues from the real axis for different relative strengths of the dissipation $\alpha$. A constant offset is expected for $\alpha = 0$ due to $\mathcal{L}_D^1$. We observe no significant scaling for clusters supported by eigenmodes of a single weight (see data points for $k \in \{1,2\}$), whereas the cluster $k = 3 \text{ to } 6$ that particularly contains adjacent weights with the same analytic center ($k=4,5$) scales approximately linear with $\alpha$. This supports the analytic results for the first-order perturbation.}
	\label{fig:fig6}
\end{figure}

Next, we move to the second-order perturbation and calculate the mean of the perturbed cluster to see the attraction phenomenon. Overall, we find the mean of the second-order perturbation in one k-local subspace is given by
\begin{align}
\overline{\lambda_{2i}}(k) &= \frac{1}{n_k} \sum_{i} \sum_{m \neq k, j} \frac{\langle x_{0i}^{(k)}|\mathcal{L}_U| x_{0j}^{(m)}\rangle \langle x_{0j}^{(m)}|\mathcal{L}_U| x_{0i}^{(k)}\rangle}{\lambda_{0}(m) - \lambda_{0}(k)}
\end{align}
where the sum does not depend on the particular basis $\{ | x_{0i}^{(k)} \rangle \}$ of a degenerate subspace containing weight $k$ Pauli strings \and $\lambda_0(k) = -2(6 k \ell - 4 k^2)/(3 \ell + 9 \ell (\ell - 1) /2)$. In the Pauli string basis, this sum reads
\begin{align}
\overline{\lambda_{2i}}(k) = \frac{1}{n_k} \sum_{i} \sum_{m \neq k, j}  \frac{(\mat{L}^{(km)}_{ij})^2}{\lambda_{0}(m) - \lambda_{0}(k)},
\label{eq:second_order_PT}
\end{align}
where $\mat{L}^{(km)}_{ij}$ are the matrix elements in the block of the matrix representation given by a $k$-local and a $m$-local Pauli string. Recall that $\mat{L}^{U}_{x,y}$ are Gaussian distributed with standard deviation $2\sigma$ and that there are $h(k,k-1) + h(k,k+1)$ non-vanishing entries in each row in the weight-$k$-block. This indicates that the mean of the perturbed cluster is given by Gaussian averaging of Eq. (\ref{eq:second_order_PT})  
\begin{align}
    \langle \overline{\lambda_{2i}}(k) \rangle = \frac{8 }{9 \ell (\ell -1)} \sum_{m \in \{k \pm 1\}} \frac{h(k, m)}{\lambda_0(m) - \lambda_0(k)},
\label{eq:second_order_PT_averaged}
\end{align}
and reduces to $\langle \overline{\lambda_{2i}}(k) \rangle \to 1/\ell^{2}$ in the thermodynamic limit. As indicated in Fig. \ref{fig:fig6}, the analytic prediction of the mean agrees well with exact diagonalization results.

\subsection{Weak dissipation}
The limit of a strong unitary contribution $\alpha \to \infty$ can be considered as a weak dissipation limit, when the Liouvillian and thus the spectrum are simultaneously rescaled with the factor $1/\alpha$, i.e. $\tilde{\mathcal{L}} = \mathcal{L}_U + \mathcal{L}_D/\alpha$. We refer to the limit $\alpha \to \infty$ also as a weak dissipation limit. In Appendix \ref{appendix:weak_dissipation}, we show the spectral transition for the actual limit $\beta \to 0$ for $\mathcal{L}_U + \beta \mathcal{L}_D$.

In the limit of weak dissipation, the spread of the eigenvalues is only determined by the Hamiltonian. Since the eigenvalues of $\mathcal{L}_U$ are given by $\lambda_U = i (E_n - E_m)$, where $E_i$ are the eigenvalues of $H$, we can calculate the standard deviation $\mathrm{std}(\mathrm{Im}(\lambda_U)) = \sqrt{2}$ using the normalization of the Hamiltonian $\mathrm{Tr}(H^2)=N$. The numerical results shown in Fig. \ref{fig:fig6} agree with the linear scaling ($\alpha \sqrt{2}$) of the imaginary extent of the spectrum of $\mathcal{L}_D + \alpha \mathcal{L}_U$. 

Furthermore, for a dominant unitary part, we observe the spectrum converging to a concentrated eigenvalue cluster around $-1$ \cite{denisov_universal_2019}, but also a single persistent decay mode supported by weight-2 Pauli strings, as in Fig. \ref{fig:spectrum_random}(h). Since the Pauli strings span the full $4^{\ell}$ dimensional Liouville operator space, one can have a decay mode equal to the Hamiltonian, which yields a trivial commutator $\mathcal{L}_U(H) = -i [H,H] = 0$. The weight-2-only nature of Hamiltonian makes this persistent decay mode share the same locality and lives in the $2$-local subspace of $\mathcal{L}_D^{0}$. Consequently, it is an eigenvector of $\mathcal{L}_D^{0}$ and $\mathcal{L}_D^{0}+\alpha \mathcal{L}_U$ with eigenvalue $\lambda_0{(2)}$. Taking $\mathcal{L}_D^{1}$ as a small perturbation approximates the persistent decay mode $\rho \approx H$ of the full Liouvillian $\mathcal{L}$ with eigenvalue $\lambda \approx \lambda_0{(2)}$.

\section{Application: Heisenberg chain}\begin{figure}[h]
    \centering
    \includegraphics[width=1\columnwidth]{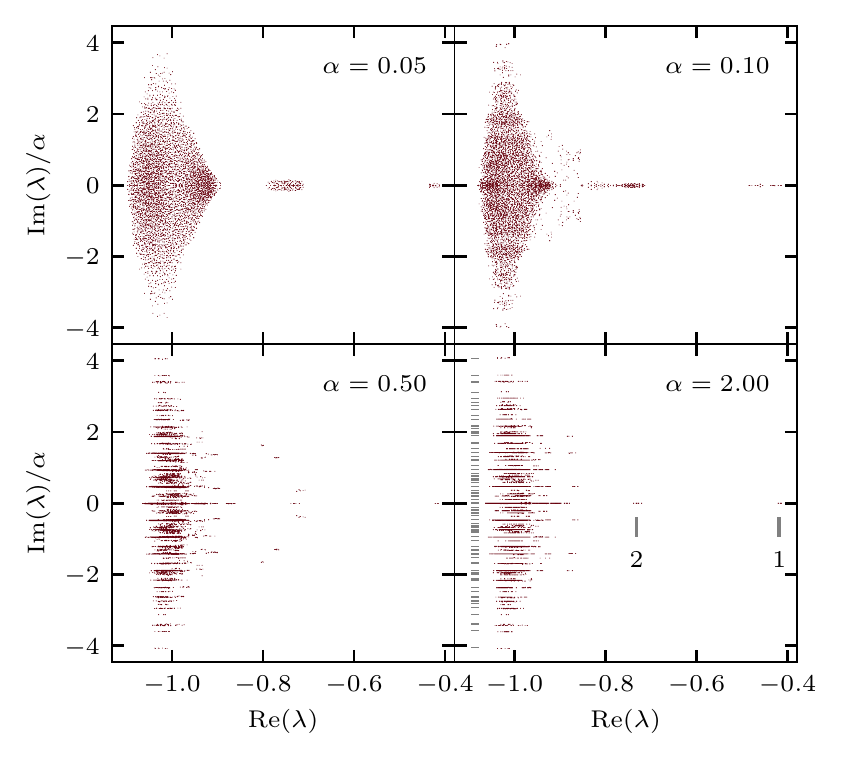}
    \caption{Spectrum of the Lindbladian for the 1D XXX Heisenberg chain with up-to-2-local random dissipation from a single realization. The upper panels demonstrate the attractions of eigenvalue clusters. The lower panel indicates the persistent clusters in the opposite limit. The left horizontal lines in the weak dissipation limit (bottom right) show the spectrum of the purely unitary Liouvillian $\mathcal{L}_U$. The vertical lines with numbers show the expected positions of persistent eigenmodes of the respective weights.}
    \label{fig:spectrum_Heisenberg_PBC}
\end{figure}

As a concrete example, we consider the XXX antiferromagnetic Heisenberg Hamiltonian 
\begin{align}
    H 
    = J \sum_{i=1}^{\ell} \left( \sigma^{i}_x \otimes \sigma^{i+1}_x +\sigma^{i}_y \otimes \sigma^{i+1}_y+\sigma^{i}_z \otimes \sigma^{i+1}_z \right),
\end{align}
where $J = 1/\sqrt{3 \ell} > 0$ describes the coupling of two neighboring spins and implies the normalization  $\mathrm{Tr}(H^2)=N$. We further use periodic boundary conditions (PBC) with random 1- and 2-local dissipation. Figure \ref{fig:spectrum_Heisenberg_PBC} shows the Lindbladian spectrum of such a 1D random Heisenberg chain for different strengths of the unitary component. When the unitary strength $\alpha$ is weak, the clusters show a similar attraction and merging behavior to the totally random 2-local Hamiltonian.

In the weakly dissipative case (large $\alpha$), we observe the eigenvalues ordered into lines with the same imaginary part due to the discrete spectrum nature of the Heisenberg Hamiltonian and remaining well-separated clusters. The emergence of the extra persistent eigenmodes is a manifestation of the symmetry properties of the Hamiltonian. The XXX spin chain has a SU(2) symmetry algebra, which means the total spin operators 
\begin{align*}
	\rho_\alpha^{(1)} = \sum_{i=1}^{\ell} \sigma^{i}_\alpha, \quad \alpha \in \{x,y,z\}
\end{align*}
satisfy $[H,\rho_\alpha^{(1)}]=0$. We note that $\rho_\alpha^{(1)}$ are sums of Pauli strings of weight 1. This local symmetry of the XXX chain implies the existence of 3 persistent eigenmodes of weight 1 with an eigenvalue that is independent of the unitary component of $\mathcal{L}$. 

Furthermore, we also find the following 7 eigenmodes of weight 2 that commute with the Hamiltonian
\begin{align*}
	\rho_1^{(2)} & = H; \quad \rho_{\alpha\beta}^{(2)} = \sum_{i \neq j} \sigma^{i}_\alpha \otimes \sigma^{j}_\beta, \quad \alpha, \beta \in \{x,y,z\},
\end{align*}
to make up 10 persistent eigenmodes in total. Other conserved operators of higher weight may be observed as separated eigenvalues at the respective positions but flow to the cluster with a real part $-1$ in the presented case.

\begin{figure}[h]
    \centering
    \includegraphics{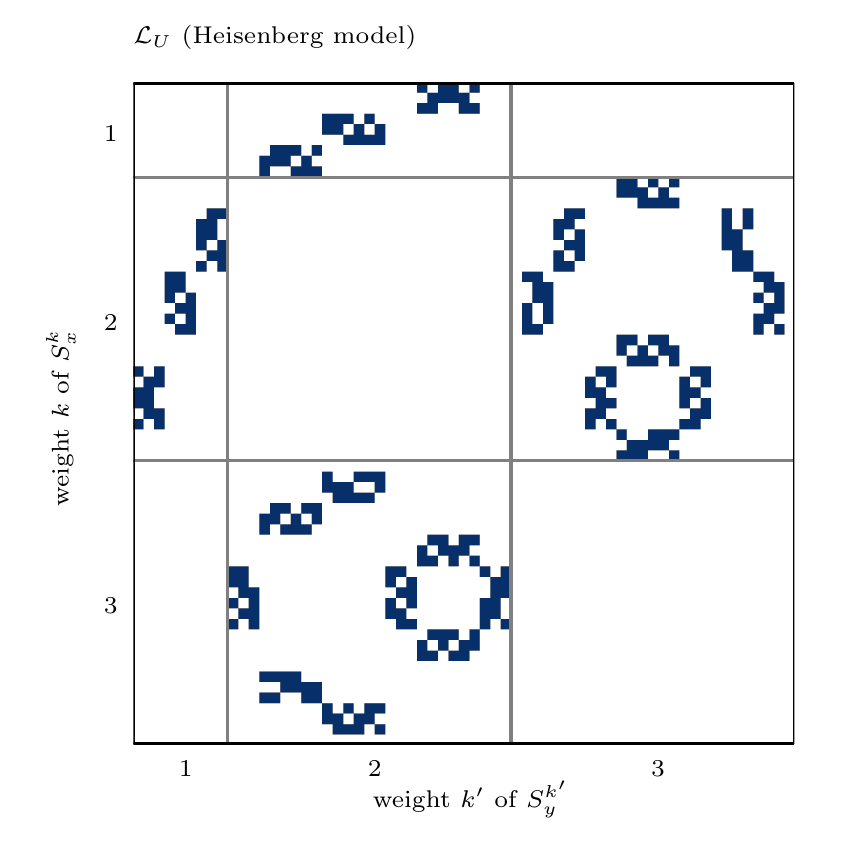}
    \caption{Single realization of the supermatrix elements of a 2-local Heisenberg $\mathcal{L}_{U}$ ($\ell = 3$) in the Pauli String basis. The XXX chain only consists of Pauli strings of weight 2, hence it only couples the adjacent weight sectors of $\mathcal{L}^{0}_{D}$, as predicted by Eq. (\ref{eq:LU_PT}).}
    \label{fig:matrix_L_U_heisenberg}
\end{figure}

The terms of the XXX chain are a subset of the totally random 2-local $\mathcal{L}_U$. Thus, the same perturbative treatment holds and leads to a sparse unitary matrix Eq. (\ref{eq:LU_PT}). Figure \ref{fig:matrix_L_U_heisenberg} shows the realization of the supermatrix elements of a 2-local Heisenberg $\mat{L}_U$ in the Pauli string basis. The weight 2 nature of the XXX chain only couples the Pauli string of adjacent weights. 

\section{Discussion and Conclusion}
In this work, we present a perturbative study on how local Hamiltonians modify the hierarchy of relaxation timescales in $k$-local strongly dissipative Liouvillians. The modified timescales are determined by the weight of the Hamiltonian terms in the Pauli string basis via coupling different weight sectors of the Liouvillian. 

For a $k$-local unitary with $k$ even, the separation of the eigenvalue clusters, which physically correspond to the hierarchy of relaxation timescales, is stable up to second-order in the strength $\alpha$ of the unitary contribution; the imaginary parts of the eigenvalues depend linearly on $\alpha$ as a first-order perturbation effect, but the real parts (responsible for the separation of eigenvalue clusters) change only in second order in $\alpha$. These results lend a certain robustness to the hierarchy of timescales stemming from the purely dissipative limit \cite{wang_hierarchy_2020} and explain why this hierarchy can be observed experimentally even though a weak unitary contribution cannot be ruled out in experiments \cite{sommer_many-body_2021}.

For a random 2-local Hamiltonian, we find the second-order perturbation changes of the average eigenvalues in each eigenvalue cluster to behave like $\left( \alpha / \ell\right)^{2}$ in the thermodynamic limit. Our perturbation analysis is applicable to the non-random XXX antiferromagnetic Heisenberg chain, where we find one persistent eigenmode in the weak dissipation limit, which we identify as the Hamiltonian. In addition to this Hamiltonian mode, we find further persistent modes in the XXX chain given by its SU(2) operators. 

Overall, our result for random Lindbladians captures the relaxation dynamics of $k$-local observables in generic open quantum systems and directly extends to non-random models as considered here.

\begin{acknowledgments}
We are grateful to one of the anonymous referees for helpful comments on jump operator normality. This project was supported by the Deutsche Forschungsgemeinschaft (DFG) through the 
cluster of excellence ML4Q (EXC 2004, project-id 390534769). 
We further acknowledge support from the QuantERA II Programme that has received funding from 
the European Union’s Horizon 2020 research innovation programme (GA 101017733), and
from the Deutsche Forschungsgemeinschaft through the project DQUANT (project-id 499347025).
\end{acknowledgments}

\bibliography{refs.bib}

\newpage
\appendix

\onecolumngrid
\section{Spectral Density }
\label{appendix:statistics}
\begin{center}
	\begin{figure*}[!h]
		\centering
		\includegraphics[width=\textwidth]{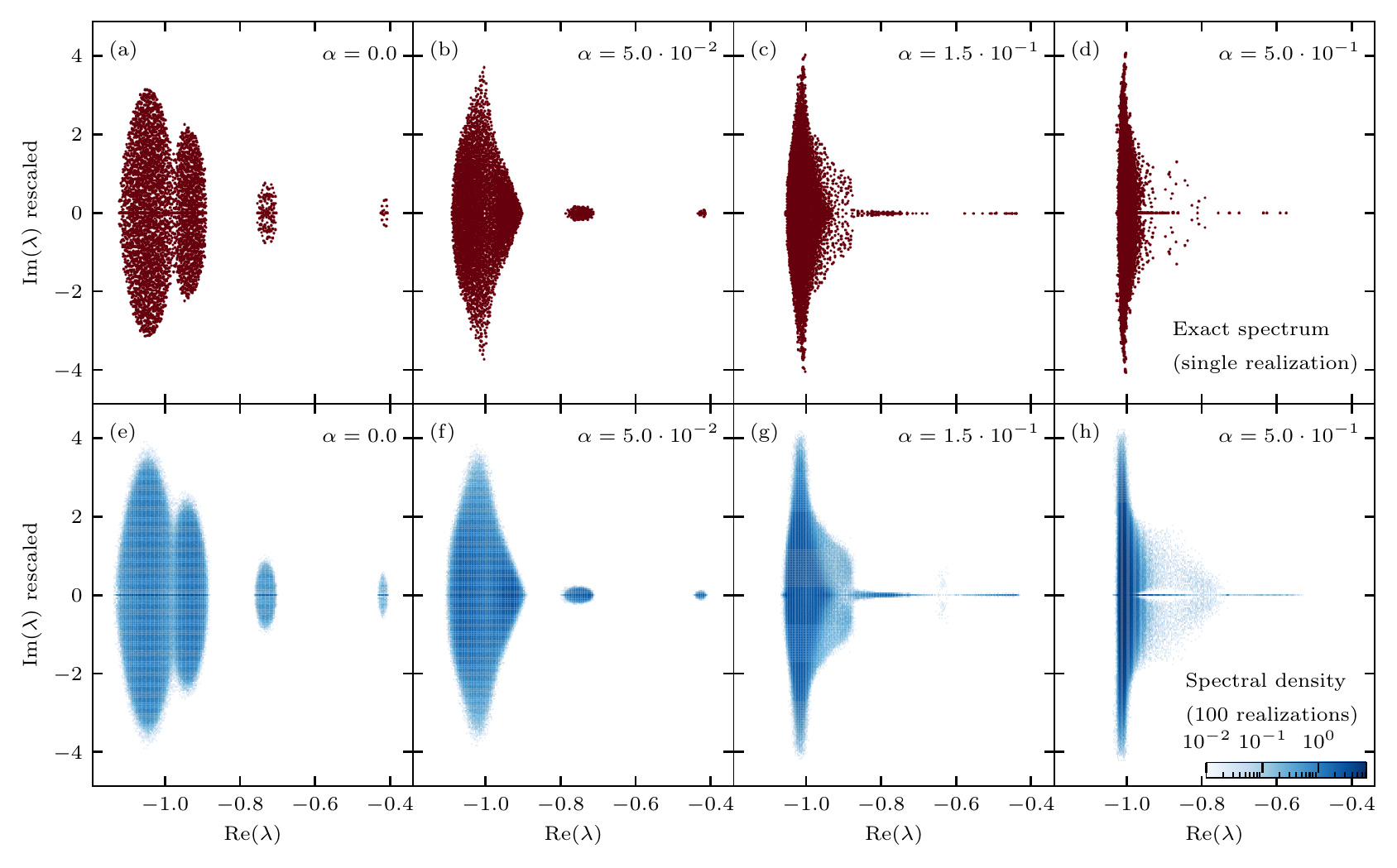}
		\caption{(a)--(d) Eigenvalue spectrum of one realization of the random Liouvillian for a system size $\ell = 6$, with weight-2 $\mathcal{L}_{U}$ and up-to-weight-2 ($k\leq 2$) $\mathcal{L}_{D}$ at 4 different relative strengths $\alpha$ of the unitary component. 
		(e)--(h) Spectral density of the same spectra averaged over 100 realizations. Note that the imaginary axis was rescales by the factor $1/\alpha$ for (b)--(d), (f)--(h), and for (a),(e) by the factor $100$. The spectral density applies for the rescaled axes.}
		\label{fig:appendix_statistical}
	\end{figure*}
\end{center}

The emergence of sharply bounded spectra which show generic structures independent from the individual (random) realization from the ensemble was already observed for various random matrices, including Ginibre matrices \cite{ginibre_statistical_1965} or a non-local random Liouvillian \cite{denisov_universal_2019}. The $k$-local random model considered in this work shows a similar self-averaging phenomenon--the spectral density of a single realization is generic for the whole random ensemble. Fig. \ref{fig:appendix_statistical} shows the spectral density of single and 100 random realizations of $\mathcal{L}$ for different unitary contributions. 

\section{Weak dissipation limit}
\label{appendix:weak_dissipation}
\begin{center}
	\begin{figure*}[!h]
		\centering
		\includegraphics[width=\textwidth]{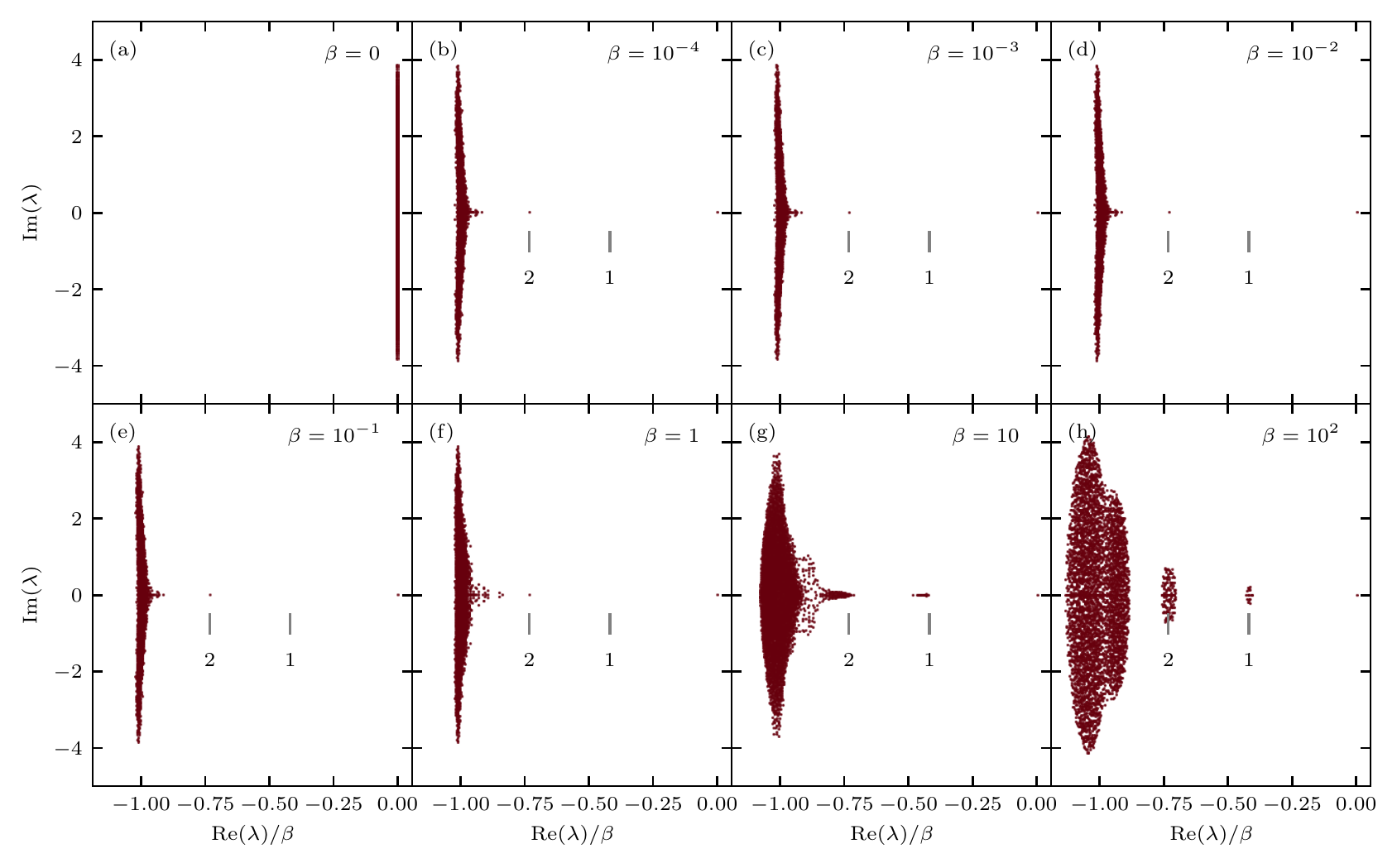}
		\caption{Eigenvalue spectrum of one realization of the random Liouvillian for a system size $\ell = 6$, with weight-2 $\mathcal{L}_{U}$ and up-to-weight-2 ($k\leq 2$) $\mathcal{L}_{D}$ at eight different relative dissipation strengths $\beta$ ($\mathcal{L} = \mathcal{L}_U + \beta \mathcal{L}_D$). 
			Note that we rescale the real part of the eigenvalues by a factor of $1/\beta$ to put these panels on the same scale. The gray lines below the spectrum give the analytic predictions for the cluster centers corresponding to weight 1 and 2 eigenstates using $\mathcal{L}^{0}_{D}$ \cite{wang_hierarchy_2020}.
			(a) Purely unitary case ($\beta = 0$), the eigenvalues are purely imaginary $\lambda = \pm i (E_n - E_m)$, where $E_n$ are eigenvalues of $H$
			(b)--(e) A non-vanishing dissipation leads to a shift of the spectrum in the real direction by $-\beta$. The shape of the spectrum does not depend on the dissipation strength for small dissipation.
			(f),(g) With increasing dissipation strength, the eigenvalue start to arrange in clusters.
			(h) Strong dissipative system with expected eigenvalue clusters.}
		\label{fig:spectrum_random2}
	\end{figure*}
\end{center}

In the main text, we have considered the spectral support for various unitary contributions from the purely dissipative case. In this section, we consider the spectrum starting from the purely unitary case to analyze the weak dissipation limit. Fig. \ref{fig:spectrum_random2} shows the clusterization of complex eigenvalues starting from the imaginary axis. 

\end{document}